\documentclass[12pt,english]{article}
\usepackage[T1]{fontenc}
\usepackage[latin1]{inputenc}
\usepackage{babel}
\usepackage{color}
\usepackage{varioref}

\makeatletter

\providecommand{\LyX}{L\kern-.1667em\lower.25em\hbox{Y}\kern-.125emX\@}

\makeatother
\begin{document}
\baselineskip .3in 

{\centering \textbf{\textcolor{black}{\LARGE Distributions of money
in model markets of economy}}\LARGE \par}
\bigskip{}

{\centering \vskip .2in \textbf{\textcolor{black}{Anirban Chakraborti}}%
\footnote{\textcolor{black}{\emph{email address}}\textcolor{black}{: anirban@cmp.saha.ernet.in}
}\par}

{\centering \textit{\textcolor{black}{Saha Institute of Nuclear Physics}}\textcolor{black}{,}
\textit{\textcolor{black}{1/AF Bidhan Nagar, Kolkata 700 064, India.}}\par}

\noindent \textcolor{black}{\vskip .3in }

{\noindent \centering \textbf{\textcolor{black}{Abstract}}\par}

\noindent \textcolor{black}{We study the distributions of money in
a simple closed economic system for different types of monetary transactions.
We know that for arbitrary and random sharing but locally conserving
money transactions, the money distribution goes to the Gibb's distribution
of statistical mechanics. We then consider the effects of savings,
etc. and see how the distribution changes. We also propose a new model
where the agents invest equal amounts of money in each transaction.
We find that for short time-periods, the money distribution obeys
a power-law with an exponent very close to unity, and has an exponential
tail; after a very long time, this distribution collapses and the
entire amount of money goes to a tiny fraction of the population.}

\textcolor{black}{\vskip 1in }


\newpage

\section{\noindent \textcolor{black}{Introduction}}

Economics deals with the real life around us. The area of economics
is not only restricted to the marketplace but also covers almost everything
from the environment to family life! Economics is the study of how
societies can use the scarce resources efficiently to produce valuable
commodities and distribute them among different people or economic
agents \cite{Sam,Key}. Financial markets exhibit several properties
that characterize complex systems. For financial markets, the governing
rules are rather stable and the time evolution of the system can be
continuously monitored. Recently, an increasing number of physicists
have made attempts to analyse and model financial markets, and in
general, economic systems \cite{Ecophy,Ecophy1}. The physics community
took the first interest in financial and economic systems, when Majorana
wrote a pioneering paper \cite{Maj} on the essential analogy between
statistical laws in physics and in the social sciences. This off-the-track
outlook did not create much interest until recent times. In fact,
prior to the 1990's, a very few professional physicists like Kadanoff
\cite{Kad} and Montroll \cite{Mon}, took much interest for research
in social or economic systems. Since 1990, physicists started turning
to this interdisciplinary subject, and their research activity is
complementary to the most traditional approaches of finance and mathematical
finance. It may be surprising to the students of physical sciences
but the first use of power-law distribution was made by an Italian
social economist Pareto, a century ago, who investigated the wealth
of individuals in a stable economy by modelling them using the distribution\[
y\sim x^{-\upsilon },\]
where \( y \) is the number of people having income greater than
or equal to \( x \) and \( \upsilon  \) is an exponent which he
estimated to be \( 1.5 \) \cite{VP}. Almost during the same time,
the first formalization of a random walk was made by a French mathematician
Bachelier in his doctoral thesis \cite{LB}, where he used the increments
of Brownian motion to model `absolute' price changes. A major part
of the recent efforts made by physicists has gone in investigating
the nature of fluctuations and their distributions in the stock markets
\cite{Ecophy}. We believe that a thorough understanding of the statistical
mechanics of the money market, especially studying of the distribution
functions, is essential. There has been some very interesting papers
along this line \cite{LS,ACE1,DY,ACE2,Sou}. \textcolor{black}{Here,
we make a very brief review of the earlier models and then propose
a new variant where the agents invest equal amounts of money in each
transaction. We find that for short time-periods, the money distribution
obeys a power-law with an exponent very close to unity, and has an
exponential tail; after a very long time, this distribution collapses
and the entire amount of money goes to a tiny fraction of the population.}

\section{\noindent \textcolor{black}{Brief review of models and distributions
of money}}

We consider a model of a closed economic system where the total amount
of money \( M \) is conserved and the number of economic agents \( N \)
is fixed. Each economic agent \( i \), which may be an individual
or a corporate, possesses money \( m_{i} \). An economic agent can
exchange money with any other agent through some trade, keeping the
total amount of money of both the agents conserved. We assume that
an agent's money must always be non-negative and therefore no debt
is permitted.

\subsection{\noindent \textcolor{black}{Random transactions\label{dy}}}

Let an arbitrary pair of agents \( i \) and \( j \) get engaged
in a trade so that their money \( m_{i} \) and \( m_{j} \) change
by amounts \( \Delta m_{i} \) and \( \Delta m_{j} \) to become \( m_{i}^{\prime } \)
and \( m_{j}^{\prime } \), where \( \Delta m_{i} \) is a random
fraction of \( (m_{i}+m_{j}) \) and \( \Delta m_{j} \) is the rest
of it, so that conservation of the total money in each trade is ensured.
The money distribution goes to the equilibrium Gibb's distribution
\cite{Stat} of statistical mechanics: \( P(m)=(1/T)\exp (-m/T) \)
where {}``temperature'' \( T=M/N \), the average money per agent
in the market, satisfying \( P(m_{i})P(m_{j})=P(m_{i}+m_{j}) \). 

Extensive numerical simulations show that this and various modifications
of trade, like multi-agent transactions, etc., all lead to the robust
Gibb's distribution {[}see Fig. 1{]}, independent of the initial distribution
the market starts with \cite{DY}. So, most of the agents end-up in
this market with very little money.

\subsection{\textcolor{black}{Transactions with constant saving\label{cs}}}

Let us introduce the concept of {}``saving'', which is a very natural
and important ingredient in economics, in our model. We assume that
each economic agent saves a constant amount of money \( m_{o} \)
before trading. Let us now consider that an arbitrary pair of agents
\( i \) and \( j \) get engaged in a trade so that their money \( m_{i} \)
and \( m_{j} \) change by amounts \( \Delta m_{i} \) and \( \Delta m_{j} \)
to become \( m_{i}^{\prime } \) and \( m_{j}^{\prime } \) ; \( \Delta m_{i}=\epsilon (m_{i}+m_{j}-2m_{o}) \)
and \( \Delta m_{j}=(1-\epsilon )(m_{i}+m_{j}-2m_{o}) \), where \( \epsilon  \)
is a random number between zero and unity, and \( m_{i}^{\prime }=m_{o}+\Delta m_{i} \)
and \( m_{j}^{\prime }=m_{o}+\Delta m_{j} \) after the trade. In
this case, the lower limit of money that an agent is allowed to possess
actually changes from zero to \( m_{o} \). Conservation of the total
money in each trade is ensured, as earlier.

The probability distribution of money still remains the Gibb's distribution
{[}see Fig. 2{]} but the money corresponding to maximum probability
\( m_{p} \), shifts from zero (or very little money) to \( m_{o} \),
and the {}``temperature'' \( T \) changes. When \( m_{o}=0 \),
we get back the case of \vref{dy}.

\subsection{\noindent \textcolor{black}{Transactions with fractional saving (marginal
propensity of saving)\label{cc}}}

\noindent We now consider the case where each economic agent saves
a fraction \( \lambda  \) of its money \( m_{i} \) before trading
\cite{ACE1}. This constant fraction of saving \( \lambda  \) is
called the {}``marginal propensity of saving'' and is a very important
quantity in economics.

Here, we choose randomly two agents \( i \) and \( j \) having money
\( m_{i} \) and \( m_{j} \), respectively. Then \( \Delta m_{i}=\epsilon (1-\lambda )(m_{i}+m_{j}) \)
and \( \Delta m_{j}=(1-\epsilon )(1-\lambda )(m_{i}+m_{j}) \), where
\( \epsilon  \) is a random number between zero and unity. Then \( m_{i}^{\prime }=\lambda m_{i}+\Delta m_{i} \)
and \( m_{j}^{\prime }=\lambda m_{j}+\Delta m_{j} \) after the trade.
Conservation of the total money in each trade is ensured, as earlier.

The results for the equilibrium distribution \( P(m) \) are shown
in Fig. 3, for some values of \( \lambda  \). The real money exchanged
randomly in any trade is less than the total money, because of the
saving by each agent. This destroys the multiplicative property of
the distribution \( P(m) \) (seen earlier for \( \lambda =0 \))
and \( P(m) \) changes from the Gibb's form to the asymmetric Gaussian-like
form as soon as a finite \( \lambda  \) is introduced. The \( \lambda =0 \)
case, same as in \vref{dy}, was practically a random-noise dominated
one and therefore effectively a non-interacting market. Introduction
of a finite amount of saving (\( \lambda \neq 0 \)), dictated by
individual self-interest, immediately makes the money dynamics cooperative
and the global ordering (in the distribution) is achieved. 

An important feature of this humped distribution \( P(m) \) at any
non-vanishing \( \lambda  \) is the variation of the most probable
money \( m_{p}(\lambda ) \) ( where \( P(m) \) becomes maximum)
of the agents \cite{ACE1}. We have, \( m_{p}=0 \) for \( \lambda =0 \)
(Gibb's distribution) and most of the economic agents in the market
end-up losing most of their money. However, even with pure self-interest
of each agent for saving a factor \( \lambda  \) of \textit{its}
own money at any trade, a global feature emerges: the entire market
ends-up with a most-probable money \( m_{p}(\lambda ) \). This \( m_{p}(\lambda ) \)
shifts in an interesting manner from \( m_{p}=0 \) (for \( \lambda =0 \))
to \( m_{p}\rightarrow T \) (for \( \lambda \rightarrow 1 \)). The
half-width \( \Delta m_{p} \) and the peak height \( P_{m_{p}} \)
of the equilibrium distribution scale practically as \( (1-\lambda )^{1/2} \)
and \( (1-\lambda )^{-1/2} \) respectively. We also note that each
individual's money \( m_{i} \) fluctuates randomly. Since the total
money is conserved, \( \langle m_{i}\rangle  \) remains constant
(\( =T \)) here, while \( \Delta m_{i} \) goes down with \( \lambda  \)
as \( (1-\lambda ) \). This is because at any time the agents keep
a fixed fraction of their individual money and receive a random fraction
of the money traded proportional to \( (1-\lambda ) \).

\section{\noindent \textcolor{black}{Model and simulation results\label{ac}}}

We now introduce a new model where we assume that both the economic
agents invest the same amount of money \( m_{min} \), the minimum
money between the agents. We choose randomly two agents \( i \) and
\( j \) having money \( m_{i} \) and \( m_{j} \), respectively.
Thus \( 2m_{min} \) is the real money which is available in the market
for random sharing. Then \( \Delta m_{i}=\epsilon 2m_{min} \) and
\( \Delta m_{j}=(1-\epsilon )2m_{min} \), where \( \epsilon  \)
is a random number between zero and unity. Then \( m_{i}^{\prime }=(m_{i}-m_{min})+\Delta m_{i} \)
and \( m_{j}^{\prime }=(m_{j}-m_{min})+\Delta m_{j} \) after the
trade. Conservation of the total money in each trade is ensured, as
earlier. Note that we may rewrite the amounts of money after trade
as \( m_{i}^{\prime }=m_{i}+\alpha m_{min} \) and \( m_{j}^{\prime }=m_{j}-\alpha m_{min} \),
where \( \alpha  \)(\( =2\epsilon -1 \)) is a random fraction whose
absolute value is less than unity, i.e., \( -1<\alpha <1 \) and \( m_{min} \)
is either \( m_{i} \) or \( m_{j} \), whichever is less. We consider
one transaction between any arbitrary pair of agents as a unit of
`time', \( t \). We made computer simulations for agents \( N=1000 \)
and the total money in the market \( M=1000 \) and then took averages
over \( 5000 \) different configurations. We studied the money distribution
\( P(m) \) at different times.

The money distribution \( P(m) \) for time \( t=10000 \) is shown
in Fig. 4. We find that the distribution obeys a power-law: \( P(m)\sim m^{-\upsilon } \),
where \( \upsilon  \) is an exponent, and has a tail which falls
off exponentially (\( \sim \exp (-\alpha m) \)). The numerically
fitted curves (indicated by the solid line and the dashed curve in
the figure) give the following exponents: \( \upsilon =0.9\pm 0.01 \),
which is very close to unity, and \( \alpha =0.25\pm 0.02 \). The
errors are obtained by eye-estimation.

We note that once an agent loses all its money, it is unable to trade
any more because \( m_{min} \) becomes zero and no other agent will
invest money for trade with this agent. Thus, a trader is effectively
driven out of the market once it loses all its money. In this way,
after an infinite number of transactions have taken place, one would
expect that only one trader survives in the market with the entire
amount of money and the rest of the traders have zero money. In our
numerical simulations, we found that for \( t=15000000 \), more than
\( 99\% \) of the traders have zero money and the rest have the entire
money of the market. This can be prevented, for example, by government
intervention via taxes, but we do not consider any such thing in our
model.

\vskip 1in

\noindent \textbf{\textcolor{black}{\large Acknowledgements}}\textcolor{black}{\large : }{\large \par}

\noindent \textcolor{black}{The author is grateful to B. K. Chakrabarti,
T. Lux and S. Pradhan for their useful comments and discussions, and
the referee for suitable corrections and suggestions.}

\vskip 1in

\noindent \textbf{Figure captions}

\noindent \vskip .1 in 

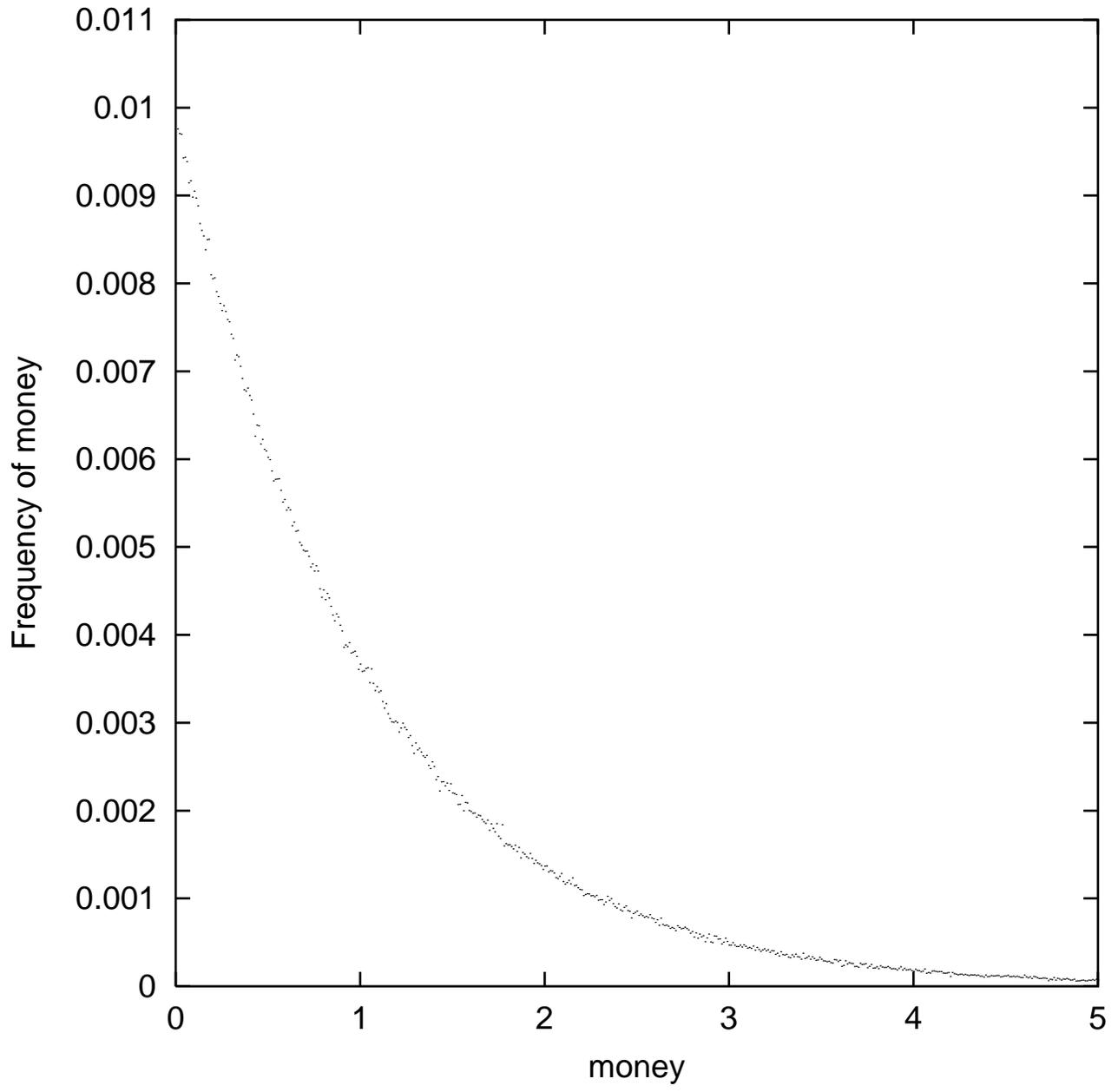
\noindent \textbf{Fig. 1} : Histogram of money, obtained from computer
simulations made for agents \( N=1000 \) and total money \( M=1000 \).
Note that in the figure, the money has been scaled by the average
money in the market.

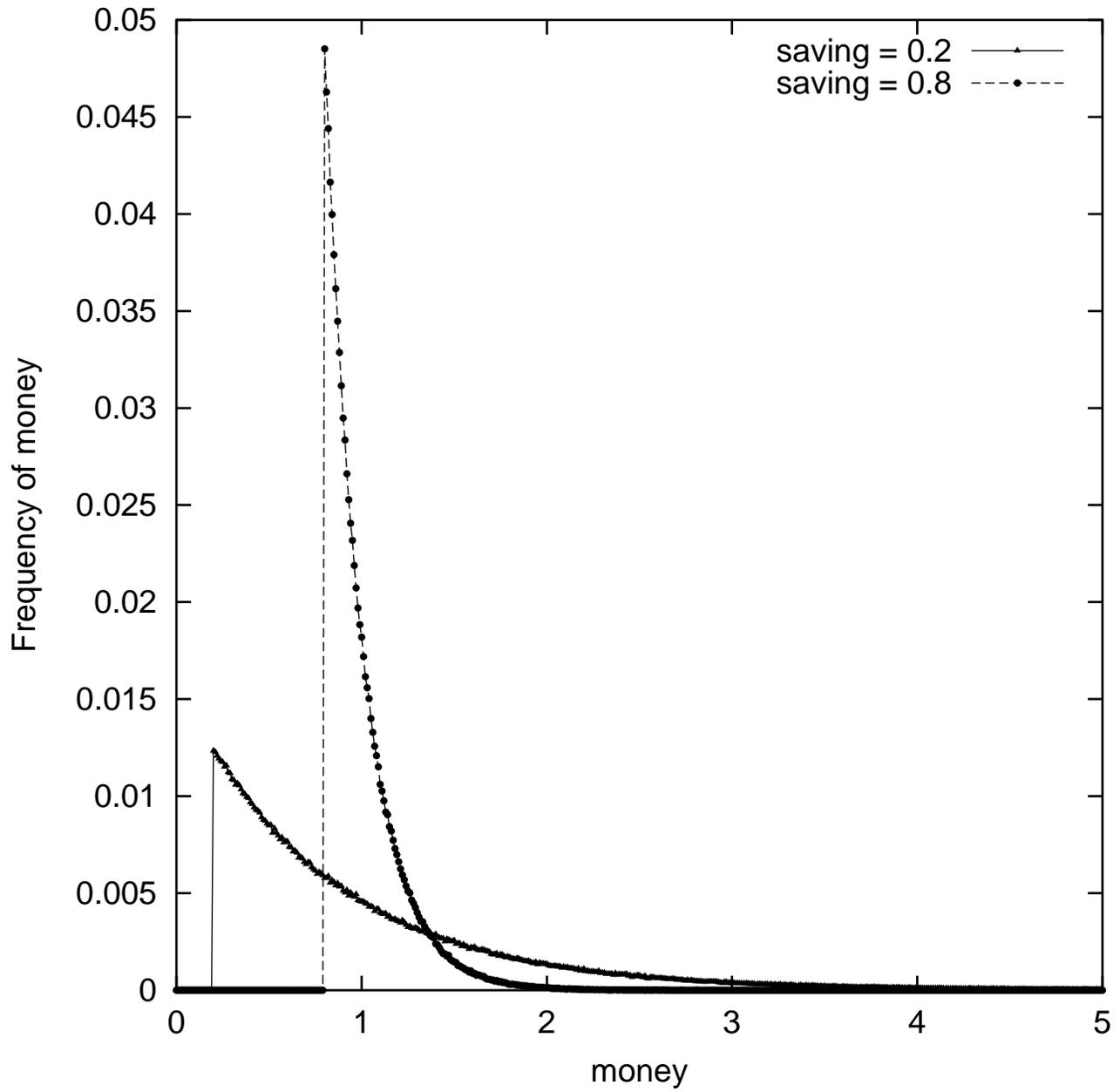
\noindent \textbf{Fig. 2} : Histogram of money for different saving
amounts \( m_{o}=0.2 \) and \( m_{o}=0.8 \), obtained from computer
simulations made for agents \( N=1000 \) and total money \( M=1000 \).
Note that in the figure, the money has been scaled by the average
money in the market.

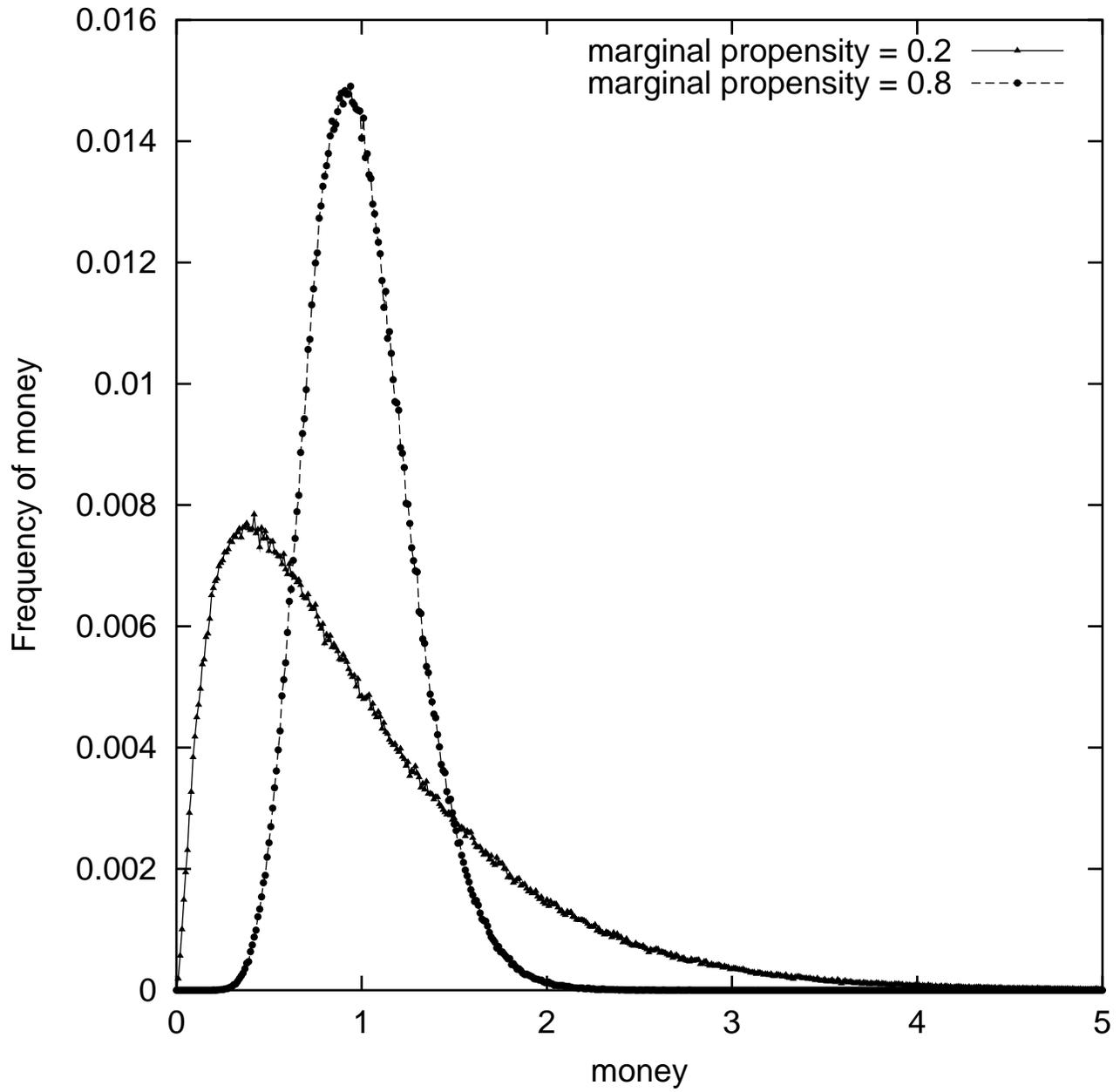
\noindent \textbf{Fig. 3} : Histogram of money for different saving
propensity factor \( \lambda =0.2 \) and \( \lambda =0.8 \), obtained
from computer simulations made for agents \( N=1000 \) and total
money \( M=1000 \). Note that in the figure, the money has been scaled
by the average money in the market.

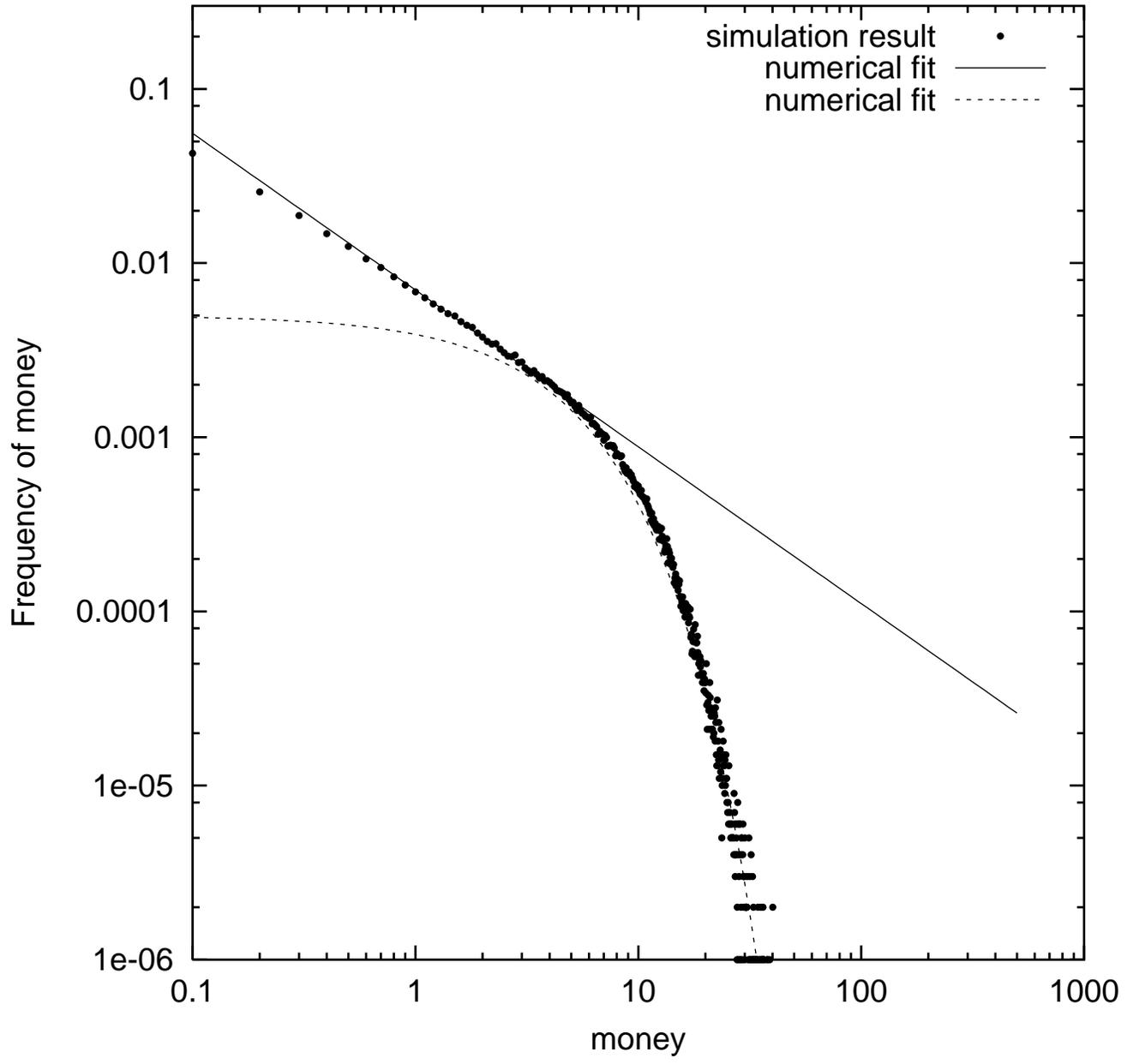
\noindent \textbf{Fig. 4} : Histogram of money plotted in the double
logarithmic scale, obtained from computer simulations made for agents
\( N=1000 \) and total money \( M=1000 \), at time \( t=10000 \).
The solid line is the numerically fitted line with slope \( \upsilon =0.9 \)
and the dashed curve is the numerically fitted exponential curve with
an exponent \( \alpha =0.25 \). Note that in the figure, the money
has been scaled by the average money in the market.

\end{document}